\documentclass[12pt,a4paper,twoside]{article}
\usepackage{times}
\usepackage{epsf}
\def\COMMENT#1{{\hrule\vskip 0.3cm\noindent\kern -5em\tt*** {#1}
\vskip 0.3cm\hrule}}
\let\epsilon=\varepsilon
\def\s{{\rm s\kern0.00010em}}
\def\f{{\rm f\kern0.00010em}}
\def\OO{{\cal O}}
\begin{document}
\begin{center}
{\Large{\bf Hydrodynamic Lyapunov Modes in Translation Invariant Systems}}\\[2cm]

{\large{ Jean-Pierre Eckmann and Omri Gat}}\\
D\'epartement de Physique  Th\'eorique, \\ 
Universit\'e de Gen\`eve, \\ 
32 Bld. D'Yvoy, CH-1211 Gen\`eve\\ Switzerland\\ [2cm]
\end{center}

{\medskip\noindent{\bf Abstract.\ }}
We study the implications of translation invariance on the tangent
dynamics of extended dynamical systems, within a random matrix
approximation. In a model system, we show the existence of
hydrodynamic modes in the slowly growing part of the Lyapunov
spectrum, which are analogous to the hydrodynamic modes discovered
numerically by [Dellago, Ch., Posch, H.A., Hoover, W.G.,
Phys. Rev. E{\bf53}, 1485 (1996)]. The hydrodynamic Lyapunov vectors
loose the typical random structure and exhibit instead the structure of
weakly perturbed coherent long wavelength waves. We show further that
the amplitude of the perturbations vanishes in the thermodynamic limit,
and that the associated Lyapunov exponents are universal.

\bigskip
\section{Introduction}\label{sec:intro}
The Lyapunov spectrum is recognized as an important diagnostic of
chaotic dynamical systems. As such, it has been studied intensely in
the context of extended systems \cite{ex1,ex2}. It has been observed
that in the thermodynamic limit the spectrum seems to approach a
continuous density, and some theoretical studies have focused on this
phenomenon \cite{ew}.
However, despite the large amount of available data, there is
an unsatisfactory degree of understanding of the relation
between the Lyapunov spectrum
of extended systems, and their global or collective properties.

In connection with these problems, the recent study 
of~\cite{posch} presents an interesting development (see also \cite{p2} for
further results and references). In the
context of molecular dynamics simulations, they find
hydrodynamic, {\it i.e.}, slow, long-wavelength behavior in the 
{\it tangent space} dynamics. Namely, they observe that the Lyapunov
vectors associated with the Lyapunov exponents of small absolute value
have {\em ordered, wave-like structure}, and that the exponents
themselves follow 
an ordered pattern.
Hydrodynamic behavior in {\it phase space} is of course
present in every extended system with a continuous symmetry. In the
models considered in~\cite{posch} the symmetries in question are
translation and Galilei invariance, precisely those which enable the
hydrodynamic description of a fluid in terms of the Navier-Stokes
equations~\cite{hydro}. However, it is for the first time that a similar
phenomenon is observed in tangent space.

In this paper, we study theoretically 
the slow Lyapunov modes (vectors and
exponents) of extended systems with translation invariance. We focus
attention on a simplified model which shares the essential
features with the more elaborate model of~\cite{posch}. This
simplified model is
constructed only in tangent space without an accompanying real space
dynamics, and is based on a random matrix approximation. As has been
often found before, in systems with strong chaos, qualitative
features of the Lyapunov spectrum are 
well reproduced by approximating the tangent matrices by 
independent random matrices with appropriately chosen
distributions~\cite{ex2,ew}.
We prove several statements on the slow Lyapunov modes of this model
in the thermodynamic limit, which show  that in this limit the
Lyapunov vectors and exponents are indeed well described as being
hydrodynamic. 

The basic reason for the existence of these hydrodynamic modes is
evidently the translation invariance. Its presence
dictates that the dynamics
are indifferent to a uniform
shift of all the particles (or their momenta), so that the associated
Lyapunov vectors are decoupled from the rest of the dynamics, and the
associated Lyapunov exponents vanish. 
We show that slowly growing large wavelength disturbances are
nearly 
decoupled for the same reason, and use this property to show
how the clean wave structure is
obtained as a result of the orthogonalization procedure which involves
all the faster growing Lyapunov vectors. It should be emphasized that
the wave-like structure characterizes the Lyapunov vector {\it at any
given instant}\/ and is not an average property. Our arguments depend
essentially on the local and hyperbolic character of
the interactions, in addition to
translation invariance. 
The absence of translation invariance has been recognized to ruin the
hydrodynamic modes~\cite{dingaling}.
In translation invariant anharmonic chains, the
absence of short time hyperbolicity seems to 
ruin the hydrodynamic modes~\cite{anh}.
We present theoretical arguments for the existence of hydrodynamic modes in the
simplified model, which are complemented by 
numerical verifications.
The outline of the paper is as follows. In section~\ref{sec:model} the
hydrodynamic phenomenology is described in some more detail, the
definition of the random matrix model is presented and motivated, and
the results are stated. They are derived in
sections~\ref{sec:specta}--\ref{sec:symplec}. Section~\ref{sec:num}
is devoted to numerical studies of the hydrodynamic properties of Lyapunov
vectors.

\section{Hydrodynamic behavior in tangent space}\label{sec:model}

The systems studied in~\cite{posch} consist (among others) of a large set of 
disks moving in a two dimensional box $\Omega$ with periodic boundary
conditions (torus 
geometry), with elastic scattering. In this case,
the phase space is $4N$-dimensional where $N$ is the number of disks.
The Lyapunov vectors have $4N$ components which we label as
$$ (\delta x_n,\,\delta y_n,\,\delta p_{x,n},\,\delta p_{y,n})\qquad
1\le n\le N\ ,$$
with evident notation. To give them a geometrical meaning
the components of the Lyapunov vectors are drawn in~\cite{posch}
at the instantaneous position of
the particles which carry a given specific index. 
That is, one constructs
a vector field $\vec v(t,\vec x)$ with values in ${\bf R}^4$, which
are defined only at the instantaneous positions $\vec x_n(t)$
of the particles, for example 
$$ v_x(t,\vec x_n(t))=\delta x_n(t)\ ,$$
and similarly for the other components.

The vector fields of the slow Lyapunov vectors, defined as the
Lyapunov vectors with small corresponding Lyapunov exponents,
are very well approximated by the long wavelength
eigenmodes of a `reverse wave equation' in the domain $\Omega$:
\begin{equation} \label{phen}\partial_t^2 \vec v(t,\vec x)=
-{1\over N^2}\nabla^2 \vec v(t,\vec x)\ \end{equation}
(note the unusual sign in front of $\nabla^2$).
That is, the vectors look like long wavelength waves with, say, $n$
nodes in the $x$ direction and $m$ nodes in the $y$ direction, and the
corresponding Lyapunov exponent is proportional to 
$$
\pm{1\over N}\sqrt{\bigg({m\over L_x}\bigg)^2+\bigg({n\over L_y}\bigg)^2}~.$$
Note that the translation modes---constant Lyapunov vectors
with zero exponents, which are trivially present in any system with
translation invariance, correspond to the special case $m=n=0$.
This phenomenology was observed in simulations with
widely varying parameters, such as aspect ratio, density, and
the shape of the particles~\cite{posch,p2}.

The tangent flow of the molecular dynamics system can be written as
\begin{equation} \label{tg1}
\partial_t\pmatrix{\delta\vec x\cr\delta\vec p}=
G(t)\pmatrix{\delta\vec x\cr\delta\vec p}\ ,\end{equation}
where the components $\delta\vec x,\,\delta\vec p$ are
column vectors with $N$ entries each of which is a vector in ${\bf
R}^2$.
The quantity $G(t)$ is the action on the tangent space induced by the
flow $\Phi(t)$ of the dynamical system: If $\psi_0$ is the instantaneous
state of the system then $G(t)$ is given by $G(t)f=D\Phi(t)_{\psi_0} f$,
where $\Phi(t)(\psi_0+\epsilon f)=\Phi(t)\psi_0 +\epsilon
D\Phi(t)_{\psi_0} f +{O}(\epsilon ^2)$.
The evolution operator of eq.~(\ref{tg1}) may be written formally as
\begin{equation}\label{evol} U={\sf T}\exp\int G(t)dt\ .\end{equation}
We proceed to construct a simplified model of the tangent dynamics, by
making a series of modifications and assumptions about the nature of
$G$ and~$U$. 

We first replace the hard-core interaction with a short range
`soft' potential. 
In  that case $G$ will have a block structure of the
form
\begin{equation}\label{tgfl} G(t)=\pmatrix{0&1\cr A(t)&0}\ ,\end{equation}
where the symmetric $N\times N$ matrix $A$ depends on the
instantaneous positions of the
particles and couples only nearest neighbors. Since the interactions are
purely repulsive, the flow is hyperbolic, which implies that
$A$ should be taken positive.

At this stage one may note that if the matrix $A(t)$ in
eq.~(\ref{tgfl}) were replaced by the negative of the discrete Laplacian, the
Lyapunov spectrum of $G$ would be precisely that described in
eq.~(\ref{phen}). However, the matrices $A(t)$ are in fact generated by
chaotic dynamics, and therefore fluctuate rapidly. Furthermore, the
particles in the gas rearrange in time, so that the positions of
the non-zero elements in the matrix also evolve. In our study we
concentrate on the first feature. That is, we show how tangent dynamics of the
type~(\ref{tgfl}) result in slow hydrodynamic modes in spite of the
fluctuations in $A(t)$; the effects of particle rearrangement may in
principle dealt with similarly, but
need to be studied further.

The above discussion allows us to conclude that the matrices $A(t)$
should have non-zero elements only at those positions which are nonzero
in the discrete Laplacian. Furthermore, momentum conservation implies
that the sum of elements in any row and column of $A$ must vanish.
This specifies completely the matrix structure of $A$, and it remains
to model the time dependence of the off-diagonal 
non-zero elements of
$A$. For this we invoke the hypothesis of strong chaos \cite{ex1,ew}: The
elements of $A$ may be treated as independent random processes, with a
correlation time $\tau$ which is short with respect to other time
scales of the system. It is commonly found that this approximation
yields results which are in good qualitative agreement with those of
the actual tangent flow.
 
With this in mind we model the evolution operator 
 $U$ by a product of independent  random matrices $S_n$
\begin{equation}\label{u} U=\prod_nS_n\ ,\end{equation}
where 
\begin{equation}
S_n\sim{\sf T}\exp\int_{(n-1)\tau}^{n\tau}G(t)dt\ .\label{sn}\end{equation}
During the time interval of length $\tau$, $A$, and therefore $G$, may
be considered constant, so the simplest model for $S$ would be
$S=\pmatrix{1&\tau\cr \tau A&1}$. However it is more convenient to
correct this form by a second order term in $\tau$ in order to
preserve the symplectic property which holds for $U$. We thus arrive
at our model \cite{ex2,ew}:
\begin{equation}
 S=\pmatrix{1&\tau\cr \tau A&1+\tau^2 A}\ .
\label{types}\end{equation}
The matrices $A$ are independent, and their off-diagonal elements are
independent and identically distributed. The actual probability distribution of
the off-diagonal elements can be chosen arbitrarily, subject to the
constraint of uniform hyperbolicity, namely that the support of the
distribution is strictly negative, and bounded away from zero.

The model as defined above makes sense in any space dimension, but for
the  sake of simplicity we study it in one dimension. There it bears
similarity to the tangent dynamics of an anharmonic
chain. However, in the latter case the matrices $S$ would be elliptic
rather than hyperbolic. As we show below this is an essential
ingredient in the mechanism for hydrodynamic modes, which are not
present in the Lyapunov spectra of anharmonic
chains~\cite{anh}. Unfortunately, we have not been able to find a
model dynamics in real space whose tangent space dynamics would
resemble that generated by the matrices of type~(\ref{types}). On the
other hand our results do not use explicitly the dimensionality of the
system and seem to be generalizable to higher dimensions.

Since the individual matrices are symplectic,
the Lyapunov exponents of~(\ref{u}) come
in pairs of equal absolute value and opposite signs. Translation
and Galilei invariance imply the existence of two vanishing
exponents. We concentrate our attention on the Lyapunov exponents
$\lambda_{N-1}$ and $\lambda_{N-2}$ of smallest positive value, and
the corresponding Lyapunov vectors $v_{N-1}$ and $v_{N-2}$.

Before we go on, it is necessary to make precise what we mean by
Lyapunov vectors. As is well-known, essentially all
numerical methods for calculating the tangent space dynamics of the
the Lyapunov vectors $v_n$ are defined as follows:
One starts with an orthogonal matrix $Q$, mutiplies it from the left
with the tangent matrix (in the present case $S$) and decomposes the
result as $SQ=Q' R$, where $Q'$ is orthogonal and $R$ is upper
triangular.
This procedur is iterated to yield a sequence of $Q_t$.
The
columns of the orthogonal matrices $Q_t$
are what we will call the
Lyapunov vectors.
The reader should note that these vectors are not the ones
whose existence is
proved in the multiplicative ergodic theorem. 

We can now state the main result of this paper. 
{\bf Existence of hydrodynamic Lyapunov modes}: {\em 
As the size $N$ of the matrices tends to infinity the
exponents $\lambda_{N-1}$ and $\lambda_{N-2}$ as well as the vectors
$v_{N-1}$ and $v_{N-2}$ are asymptotic to the exponents and vectors
that would be obtained if the matrices $A$ where replaced everywhere
by the negative of the discrete Laplacian (properly rescaled).}
The statement holds for the Lyapunov vectors, which are random
objects, in probability.

The statement is spelled out only for two Lyapunov modes, which
have nearly equal exponents, and where the deviation from hydrodynamic
behavior is the smallest. However, as will become apparent from the
arguments below, the result can be extended to a number
of Lyapunov modes near the middle of the spectrum which is
proportional to $\sqrt N$. Our numerical studies also indicate that
this is in fact true.

As already explained, the basic reason for the existence of the
hydrodynamic modes is translation invariance. However, this general
observation is not sufficient, and the actual proof is not
trivial. 
It depends on the random nature of successive matrices, {\it i.e.}, on
strong chaos.
Our strategy will be to show the existence of hydrodynamic
modes first in the spectrum of a {\em single} matrix of type $A$,
i.e., a negative random Laplacian (sec.~\ref{sec:specta}); then this
will be used to show that 
such modes exist in the Lyapunov spectrum of non-symplectic products
of type $\prod(1+A_n)$ in sec.~\ref{sec:oneplusa}, which in turn will be
used to show the same property for symplectic products in
sec.~\ref{sec:symplec}.

\section{Spectral properties of a single matrix}\label{sec:specta}
The matrix $A$ defined in section~\ref{sec:model} takes in one dimension
the explicit form
\begin{equation} \label{forma}
A=\left (\matrix
{ a_1+a_2& -a_2&0&0& \cdots&0 &-a_1\cr
  -a_2 & a_2+a_3 & -a_3 &0& \cdots &0& 0\cr
  \vdots& & & &&&\vdots\cr
  -a_1& 0&0&0&\cdots&-a_N&a_N+a_1\cr
}\right )~,
\end{equation}
where the $a_n$ are positive identically distributed independent
random numbers. The distribution of the $a_n$ is arbitrary, subject to
the condition
$0<a_{\rm min}<a<a_{\rm max}$ with
$a_{\rm min}<
a_{\rm max}$, and normalized such that $\left<a^{-1}\right>=1$, for later
convenience. 
The $\left<\cdot\right>$ always denote expectation
with respect to the probability distribution of the $a$. We are not
going to assume that the width of the distribution is small.

The matrix $A$ may be written as a product
\begin{equation}\label{dad}
A=-\underline \partial{\cal A}\overline\partial\end{equation}
 where
${\underline\partial}$ and  ${\overline\partial}$ are the discrete derivatives
whose action on a vector $v\in{\bf R}_N$ is
\begin{equation} ({\underline\partial}v)_n=v_n-v_{n-1}\ ,\qquad
({\overline\partial}v)_n=v_{n+1}-v_{n}\ ,\end{equation}
and ${\cal A}$ is a diagonal matrix with diagonal elements $a_n$. (The
indices are extended periodically so that $a_{N+1}\equiv a_1$). If all
the $a_n$ were equal to one, $-A$ would reduce to the discrete
Laplacian matrix $\partial^2\equiv{\underline\partial}{\overline\partial}$.

We define the Fourier transform matrix $F$ with elements
\begin{equation} 
F_{kn}={1\over\sqrt N} e^{i{2\pi\over  N}kn}~,\end{equation}
which is a unitary transformation taking ${\bf R}^N$ to $\tilde{\bf C}^N$,
the subset of ${\bf C}_N$ (with standard basis vectors $e_n$)
consisting of vectors $\tilde v$ for which
$v_{-k}=v^*_k$, which is an $N$ dimensional vector space over ${\bf R}$.
The components of $A$ in the new basis are
\begin{equation} \tilde A_{kl}\equiv(FAF^\dagger)_{kl}=
\mu_k^*(\left<a\right>\delta_{k,l}+\tilde a_{k-l})\mu_l
\ ,\end{equation}
where $\mu_k=1-\exp\left(-{2\pi i\over N}k\right)$, and $\tilde a$ is
related to the Fourier transform of $a$ considered as a vector in 
${\bf R}^N$ by
\begin{equation} \tilde {a}=N^{-1/2}F({\vec a}-\left<{\vec a}\right>)
\ .\label{tilda}\end{equation}
The random variables $\tilde a_k$ are centered, and as sums of
independent random numbers their `single-point'
distribution is nearly Gaussian with
variance
\begin{equation} \left<|\tilde a|^2\right>={\left<a^2\right>-
\left<a\right>^2\over N}
\ ,\end{equation}
so that they are typically
 small, of order $\OO(N^{-1/2})$. The joint distribution
is not Gaussian.

Note that $\mu_0=0$, so that  row  $0$ and column  $0$ of 
$\tilde A$ are zero, with the
translation vector $e_0$ being trivially a zero eigenvector.
We define the slow subspace
\begin{equation} V_\s ={\rm Span}(\{e_0,e_1,e_{-1}\})\cap\tilde{\bf C}^N\ ,
\end{equation}
and its orthogonal complement, the fast subspace 
$V_\f $. We will consider often below the block
decomposition of $\tilde A$ and other matrices into the fast and slow
subspaces, {\it e.g.},
\begin{equation} \tilde A=\pmatrix{A_{\f\f}& A_{\f\s}\cr A_{\s\f}&A_{\s\s}}
\ .\end{equation}
Note that $V_\f $ contains slow as well as fast modes.

The block $A_{\s\s}$ has small norm of order $\OO(N^{-2})$, and the off-diagonal
blocks have norm of order $\OO(N^{-1})$. However, there are more specific
properties of $A$ which are needed to establish the existence of hydrodynamic
eigenmodes. Consider the
eigenvalue problem $Av=\lambda v$. Letting $v={\underline \partial}u$, and
using the representation~(\ref{dad}) gives an equation for $u$
\begin{equation} -\partial^2u=\lambda{\cal A}^{-1}u\ .
\label{ev}\end{equation}

It is convenient to proceed by writing eq.~(\ref{ev}) in Fourier
component form
\begin{equation}\label{evc} (|\mu_k|^2-\lambda)\tilde u_k=
\lambda\sum_q\tilde b_{k-q}\tilde u_q\ ,\end{equation}
with the $\tilde b_k$ bearing a relation to $a_n^{-1}$
analogous to that between $\tilde a_k$ and $a_n$, namely
\begin{equation} \tilde b_k=N^{-1/2}F_{kn}({1\over a_n}-1)\ .
\end{equation}
Since $\left<{1\over a^2}\right><1/a_{\rm min}^2$ is $\OO(1)$ we find that
the $\tilde b_k$ are $\OO(N^{-1/2})$ for the same reason that the 
$\tilde a_k$ are.

We claim that {\em given a fixed $m$, and for $N\rightarrow\infty$ the
system~(\ref{evc}) has two linearly independent solutions 
$u^{(\pm m)},\,\lambda_{\pm m}$ such that}
\begin{equation}\label{sca1}
{1\over |u^{(\pm m)}_m|}
\sum_{|k|\ne m}|u^{(\pm m)}_k|=\OO(N^{-1/2})\ ,\end{equation}
{\em and}
\begin{equation}\label{sca2}
\left|{\lambda_{\pm m}\over|\mu_m|^2}-1\right|=\OO(N^{-1/2})\ .
\end{equation}
We justify the claim by showing that eqs.~(\ref{sca1}) and~(\ref{sca2})
are consistent with the eigenvalue equation~(\ref{evc}).
For this we rewrite~(\ref{evc}) as
\begin{equation}\label{evc2}
 \tilde u^{(m)}_k={\lambda_m\over|\mu_k|^2-\lambda_m}
\sum_q\tilde b_{k-q}\tilde u^{(m)}_q\ .\end{equation}
We assume that~(\ref{sca1}--\ref{sca2}) hold; this implies that the sum
over $q$
in~(\ref{evc2}) is dominated by the two terms with $q=\pm m$, that is,
\begin{equation} \label{evc3}
\tilde u^{(m)}_k={|\mu_m|^2\over|\mu_k|^2-|\mu_m|^2}
(b_{k-m}\tilde u^{(m)}_m+b_{k+m}\tilde u^{(m)}_{-m})\ ,\qquad
\mbox{for $|k|\ne m$.}\end{equation}
On substituting this expression in the left-hand-side of~(\ref{sca1}) the
sum over $k$ is observed to be
local, in the sense that it is dominated by terms with $|k|\sim m$, where
$|\mu_k|^2\sim\left({2\pi k\over N}\right)^2$. Since $\tilde b_k$ is
$\OO(N^{-1/2})$, assumption~(\ref{sca1}) is
verified. On the other hand, using~(\ref{sca1}) in~(\ref{evc2}) for $k=m$
gives
\begin{equation} \tilde u^{(m)}_m={\lambda_m\over|\mu_m|^2-\lambda_m}
(b_{0}\tilde u^{(m)}_m+b_{2m}\tilde u^{(m)}_{-m})\ .\end{equation}
Since $b_0$ and $\tilde b_{2m}$ are $O(N^{-1/2})$ it follows that
${\lambda_m/(|\mu_m|^2-\lambda_m)}=O(N^{1/2})$
verifying~(\ref{sca2}),
which shows that~(\ref{sca1}--\ref{sca2}) are
indeed consistent with~(\ref{evc}).

In terms of the original variables $v$, eq.~(\ref{evc3}) reads
\begin{equation} \label{evc4}
\tilde v^{(m)}_k={\mu_k^*\mu_m\over|\mu_k|^2-|\mu_m|^2}
(b_{k-m}\tilde v^{(m)}_m+b_{k+m}\tilde v^{(m)}_{-m})\ ,\qquad
\mbox{for $|k|\ne m$,}\end{equation}
so that the norm of $v^{(m)}_\perp$, the component of $v^{(m)}$
orthogonal to $e_{\pm m}$ is small,
$$ \|v^{(m)}_\perp\|^2\sim {1\over N}\sum_{|k|\ne m}{k^2m^2\over
(k^2-m^2)^2}=O\left({1\over N}\right)\ .$$
In words, these eigenvectors are almost pure Fourier modes, {\it i.e.},
eigenvectors of the discrete Laplacian.

For further developments we also need to show these modes are the only
ones with eigenvalues of order $\OO(N^{-2})$. 

This is established easily by
noting that the sharp cutoff on the probability distribution of the
$a$ implies that every realization $A$ satisfies the bounds
\begin{equation}
\label{bounda} -a_{\rm min}\partial^2<A<-a_{\rm max}\partial^2\ ,
\end{equation}
and then by the minimax principle it follows that the $p$th eigenvalue
of $A$ is larger than $a_{\rm min}$ times the $p$th eigenvalue of
$-\partial^2$ (sorting the eigenvalues of both matrices in increasing order).

The results of this section can be summarized using the decomposition
of $\tilde{\bf C}^N$ into slow and fast subspaces defined above.
We have shown there exist small numbers $\epsilon$ and $\lambda$, and
a number $0<\alpha<1$, such that the matrix
$\tilde A$ can be block diagonalized, 
\begin{equation}\tilde A=RDR^T\ ,\qquad RR^T=1\,,\qquad D=\pmatrix{D_\f &0
\cr0&D_\s }\ ,\label{blockd}\end{equation}
with the
off-diagonal blocks bounded by $\|R_{\s\f}\|,\,\|R_{\f\s}\|<\epsilon$, and
the diagonal blocks obeying 
\begin{equation}\label{block1} D_\f >\lambda>\alpha\lambda>D_\s \ge0\ ,
\end{equation}
and furthermore
\begin{equation}\label{block2} \|A_{\s\s}\|<\alpha\lambda\ .\end{equation}

The orders of magnitude for $\lambda$ and $\epsilon$ are
$\epsilon=\OO(N^{-1/2})$ and $\lambda=\OO(N^{-2})$, whereas
$\alpha\sim1/4$. However, to keep the discussion reasonably general we
are not going to use these specific values in our arguments
below. Rather, we will make statements regarding arbitrary matrices
which satisfy the conditions~(\ref{blockd}--\ref{block2}).

Although this will not be used below, it is relevant to note that if
we let $m$ vary, the small parameter in~(\ref{sca1}) and~(\ref{sca2})
becomes $m/N^{1/2}$. This means that we can expect a number of
hydrodynamic eigenmodes which is proportional to $\sqrt N$.
Another way to see this is related to the study of the vibrations of
one-dimensional disordered lattices which are 
modeled precisely by the eigenmodes of matrices of
type~(\ref{forma}). There it is known that the localization length $\xi$ is
proportional to $\lambda_m^{-1}$. Since $\lambda_m\sim ({2\pi / N})^2$
the localization length will reach $N$ when $m=\OO(N^{1/2})$. Thus,
again, we only expect 
wave-like modes when $m<\OO(N^{1/2})$.

\section{Products of matrices of the form $1+\tau A$}\label{sec:oneplusa}
In this section we use the properties derived in
section~\ref{sec:specta} to derive the existence of hydrodynamic
modes in the
Lyapunov spectrum of the product $\prod_n (1+\tau A_n)$,
where the matrices $A_n$ are independent realizations of the random
matrix defined in eq.~(\ref{forma}). Beside providing a step towards
proving the existence of hydrodynamic modes in symplectic products,
such a product may be regarded as the discrete approximation to a
continuous tangent flow given by
\begin{equation}\label{btgfl} U={\sf T}\exp\int A(t)dt\ .\end{equation}
[Compare eqs.~(\ref{evol}) and~(\ref{tgfl}).] Although this does not
correspond to the tangent flow of a mechanical system, it is
nonetheless the simplest example where hydrodynamic Lyapunov modes can
be expected. For convenience of further analysis
we absorb $\tau$ into the definition of $A$ and change to Fourier
basis once and for all, so the problem becomes 
that of a product 
\begin{equation}\label{1plusa}\prod_n (1+\tilde A_n)\ .\end{equation}

Since the Lyapunov exponents of the slow part are expected to be
smaller than the rest we aim at showing that the first $N-3$ Lyapunov
vectors span a subspace $L_\f $ (of $\tilde{\bf C}_N$) which is almost
orthogonal to $V_\s $ in the sense that for any two unit vectors $u_\f \in
L_\f $ and $v_\s \in V_\s $ one has $|u_\f \cdot v_\s |\ll1$.
Although the subspace $L_\f $
changes after each step, 
we show below that the `almost orthogonality'
is propagated from step to
step.

To show this, we propose the following scheme. Take an arbitrary vector
$u\in L_\f $, whose components in $V_\f $ and $V_\s $ are $u_\f $ and $u_\s $
respectively, normalized so that $\|u_\f \|=1$, and assume that $u_\s $ is
small. The action of $1+A$ generates a new normalized vector $u'$ by

\begin{equation}\label{up} u'={(1+A)u\over\|[(1+A)u]_\f \|}\
,\end{equation}
where $[\cdot]_\f$ is the projection onto the $\f$-component
We would like to show that $u_\s $ remains small after repeated
iteration of this process.


The block diagonalization~(\ref{blockd}) shows that the subspaces
$V_\f $ and $V_\s $ are indeed almost invariant under the
transformation
$\tilde A$. However,
in trying to apply this fact to the Lyapunov vectors of the
product~(\ref{1plusa}) we immediately encounter the danger that the
small perturbations may accumulate. The basic problem is that a vector
in $V_\s $ is contracted with respect to the `slowest' direction in
$V_\f $ by a factor of only $1-\OO(\lambda)$ (as can be seen from the
bounds on $D_\f $ and $D_\s $), whereas the perturbations which tilt a
vector in $L_\f $ with respect to $V_\f $ are of order $\epsilon$, which is
the typical size of the off-diagonal blocks
[cf.~eqs.~(\ref{blockd})--(\ref{block2})] and since we are interested
in the case 
$\lambda\ll\epsilon$ this 
contraction is not strong enough to overcome the perturbation. 

This order of magnitude argument can be made explicit by constructing
a series of matrices with the properties given by
eqs.~(\ref{blockd})--(\ref{block2}), which take a vector in $V_\f $ and
rotate it such that the outcome is a vector which has an angle
with $V_\f $ of order 1. This counter-example is
given in appendix~\ref{ap:counter}.

An essential ingredient in the construction of this counter-example
is that the 
matrix $\tilde A$ has to be chosen specifically given $u$ which
in turn depends on former realizations, in violation of the independence
assumption. In other words, although such a `bad' sequence is possible
one naturally expects that this is an event with very low
probability. Typically the perturbations to $u_\s $ generated by the
off-diagonal part of the matrices $R$ do not have the same direction,
and should 
serve to cancel one another. Therefore the statement one can hope to
show is that {\em in the sequence generated by iteration of
eq.~(\ref{up}), the probability that $\|u_\s \|>C\epsilon$ for some fixed
$C$ is very small}, as was shown for a similar example in \cite{lsy}.
Here we will only prove the weaker statement that 
the variance $\left<\|u_\s \|^2\right>$ is $\OO(\epsilon^2)$, and take
that as an indication that the probabilistic statement is correct, since
the behavior of higher moments can be treated in an analogous manner.

To prove this statement we look at the $s$-component of
eq.~(\ref{up}), 
\begin{equation}u_\s '={A_{\s\f}u_\f +(1+A_{\s\s})u_\s \over
\|(1+A_{\f\f})u_\f +A_{\f\s}u_\s \|}\ .\end{equation}
The quantity
$\left<\|u_\s '\|^2\right>$ is a sum of three terms $E_1+E_2+E_3$:
\begin{equation} E_1=\left<{\|A_{\s\f}u_\f \|^2\over\ell^2}\right>,~
E_2=2\left<{A_{\s\f}u_\f \cdot(1+A_{\s\s})u_\s \over\ell^2}\right>,~
E_3=\left<{\|(1+A_{\s\s})u_\s \|^2\over\ell^2}\right>\ ,\end{equation}
where $\ell=\|(1+A_{\f\f})u_\f +A_{\f\s}u_\s \|$.

To bound these terms we first need a lower bound
on the denominator $\ell$. Let $v_\f =R_{\f\f}^Tu_\f +R_{\f\s}^Tu_\s $,
and define $d$ by
\begin{equation}\|D_\f v_\f \|\equiv d\|v_\f \|\ .\end{equation}
Note that $d$ can vary widely between $\OO(1)$ values and $\OO(\lambda)$.
But, using the lower bound on $D_\f $~of (\ref{block1}), we see that
\begin{equation} \|(1+D_\f )v_\f \|^2=\|v_\f \|^2+2v_\f \cdot D_\f v_\f +\|D_\f v_\f \|^2
\ge(1+2\lambda+d^2)\|v_\f \|^2\ .\label{1+d}\end{equation}
Expanding $\ell$ as
\begin{equation} \ell=\|R_{\f\f}(1+D_\f )v_\f +R_{\f\s}D_\s v_\s \|\ ,\end{equation}
and using the estimates $\|R_{\f\s}D_\s \|=\OO(\epsilon\lambda)$ and
$\|1-R_{\f\f}\|=\OO(\epsilon^2)$ (cf.~eq.~(\ref{block1})), we get
from~(\ref{1+d}) the desired lower bound on $\ell$:
\begin{equation} \ell^2>(1+2\lambda+d^2)\|v_\f \|^2(1-\OO(\epsilon))
\ .\end{equation}

We can now bound $E_1$, $E_2$ and $E_3$.
First, we have
\begin{equation} \|A_{\s\f}u_\f \|=\|R_{\s\f}D_\f v_\f\| +\OO(\lambda\epsilon)<
\epsilon d\|v_\f \|+\OO(\lambda\epsilon)\ .\end{equation}
Thus, neglecting higher order corrections in $\epsilon $, we get
\begin{equation} E_1<{\epsilon^2 d^2\over1+2\lambda+d^2}\
.\end{equation}

The bound on the term $E_2$ makes essential use of the translation
invariance. For this, we note that
\begin{equation}{A_{\f\s}u_\s \over\|(1+A_{\f\f})u_\f \|^2}
\label{e2p1}\end{equation}
transforms as a vector, that is, its $k$th component is multiplied by
$\exp(i{2\pi\over N}kx)$ under a relabeling of the coordinates
$n\rightarrow n+x$. Therefore, because of translation invariance, the
expectation value 
of~(\ref{e2p1}) must remain invariant under such transformation, which
means it must vanish.
Since the denominator in $E_2$ is $\ell^2$ (which also depends on
$u_\s $) and not $\|(1+A_{\f\f})u_\f \|^2$, we need some gymnastics to
exhibit the vanishing term.
In order to see this we
write

\begin{equation}\begin{array}{l}\displaystyle
 E_2=2\bigg<{A_{\s\f}u_\f \cdot u_\s \over\|(1+A_{\f\f})u_\f \|^2}+
{A_{\s\f}u_\f \cdot A_{\s\s}u_\s \over\|(1+A_{\f\f})u_\f \|^2}\\
\displaystyle\qquad-
{A_{\s\f}u_\f \cdot(1+A_{\s\s})u_\s \left[2(1+A_{\f\f})u_\f \cdot
A_{\f\s}u_\s +
\|A_{\f\s}u_\s \|^2\right]\over \|(1+A_{\f\f})u_\f \|^2\ell^2}\bigg>
\ .\end{array}\label{e2}\end{equation}
The first term in~(\ref{e2}) vanishes because of
translation invariance, as explained before. The
second term is bounded by
\begin{equation}2\alpha\lambda\epsilon\left<\|u_\s \|\right>\end{equation}
and the dominant part of the third is
\begin{equation} 4{\bigl(A_{\s\f}u_\f \cdot u_\s \bigr)\bigl (
(1+A_{\f\f})u_\f \cdot A_{\f\s}u_\s\bigr ) 
\over \|(1+A_{\f\f})u_\f \|^2\ell^2}<{4d\epsilon^2\left<\|u_\s \|^2\right>\over
1+2\lambda+d^2}\ .\end{equation}
The last term is bounded by
$$ E_3<{1+2\alpha\lambda\over1+2\lambda+d^2}
\left<\|u_\s \|^2\right>\ .$$
Collecting the bounds yields
\begin{equation} \left<\|u_\s '\|^2\right><{d^2\epsilon^2+
(1+2\alpha\lambda+4d\epsilon^2)\left<\|u_\s \|^2\right>\over1+2\lambda+d^2}
+2\alpha\lambda\epsilon\ .\label{usp}\end{equation}
It appears from~(\ref{usp}) that although large perturbations are
possible when $d$ is $\OO(1)$, the contraction rate increases precisely
enough to compensate this contribution. Thus if $\left<\|u_\s \|^2\right>$ is
$\OO(\epsilon^2)$ to start with, it will stay so indefinitely.

In summary, assuming that the variance is indeed a measure of typical
fluctuations, we have shown that, for $N\gg1$, the
subspace $L_\f $ spanned by the first $N-3$ Lyapunov vectors of the
product~(\ref{1plusa}) is, with
very high probability, almost
orthogonal to $V_\s $. This implies that the last three Lyapunov vectors
(including the translation) remain approximately in $V_\s $. This means
by definition that they are hydrodynamic, in the sense that they are well
approximated by eigenvectors of the discrete Laplacian. Since the
action of $\tilde A$ on $V_\s $ has two eigenvalues close to $(2\pi/N)^2$
as shown in the previous section, it follows as a corollary that
the two smallest non-trivial Lyapunov exponents have approximately
this value, so that they are also hydrodynamic. This completes the
demonstration.

\section{Products of symplectic matrices}\label{sec:symplec}
We now turn to products of matrices of the form
\begin{equation} S=\pmatrix{1&\tau\cr \tau A
&1+\tau^2 A}\ .\label{sy}
\end{equation}
We disregard the two translation modes in $S$ for convenience and view
the matrices $S$ as $(2N-2)\times(2N-2)$ matrices.
Let us recall that since the matrices $S$ are symplectic and
hyperbolic, the Lyapunov exponents are non-zero and
come in pairs of opposite signs. We concentrate on modes number
$N-2$ and $N-1$ which are the smallest positive ones.

We reduce the problem to an equivalent one to which
the results of section \ref{sec:oneplusa} can be applied directly.
We denote by $L_+$ the subspace spanned by
the first $N-1$ Lyapunov vectors. It is spanned by a set of $N-1$
independent vectors, which we display in the form of a $(N-1)\times(2N-2)$
matrix ${\cal V}$. The $N-1$ vectors can always be chosen in such a way
that ${\cal V}$ is of the normal form
\begin{equation}\label{can} {\cal V}=\pmatrix{{\bf1}\cr V}\ ,
\end{equation}
where both blocks are $(N-1)\times(N-1)$.
Acting on ${\cal V}$ with $S$ gives a spanning set of the image
subspace $L_+'$,

\begin{equation} S{\cal V}=\pmatrix{1+\tau V\cr \tau A+
(1+\tau^2\tilde A)V}
\ ,\label{sl}\end{equation}
and changing basis to normal form gives 
${\cal V}'=\pmatrix{{\bf1}\cr V'}$ where
\begin{equation}
V'=\tau A+{V\over1+\tau V}\ .
\label{vpp}\end{equation}
A convenient property of this matrix dynamical system is that if $V$ is
symmetric to begin with, it stays so as a consequence of
the symplectic property of $S$ \cite{partovi}.

By definition, any vector $v\in L_+$ has a block representation
$v=\pmatrix{u,\\Vu}$. From eq.~(\ref{sl}) it follows that its image is
$\pmatrix{u',\\V'u'}$, where
\begin{equation} u'=(1+\tau V)u\ .\end{equation}
Hence, the first $N-1$ Lyapunov modes of
the products of the $S$
are the same as those of the
product $\prod_n(1+\tau V_n)$ where the matrices $V_n$ are evolving
according to eq.~(\ref{vpp}): $V_{n+1}=\tau A_n + V_n/(1+\tau V_n)$.

In view of this equivalence,
it suffices to show that the matrix $V$ has
the properties formulated in eqs.~(\ref{blockd}--\ref{block2}) and to
apply the results of
section~\ref{sec:oneplusa}. 
First note if $A_n=-\partial^2$ (minus the discrete Laplacian) for all
$n$,
then the $V_n$
converge to $f(-\partial^2)$, where
$$ f(x)={\tau x\over2}+\sqrt{x+\left({\tau x\over2}\right)^2}$$
is the larger root of the quadratic equation 
$f(x)=\tau x+{f(x)\over1+\tau f(x)}$.
For small $x>0$ this is close to $x^{1/2}$, and therefore
we assume that $V$ has a representation of the type given by
eqs.~(\ref{blockd}--\ref{block2}), with $\epsilon=\OO(N^{-1/2})$ as
before and $\lambda$ is now
$f(4\pi/N^2)=\OO(N^{-1})$.
The aim is to show that this property
is carried on to $V'$.

In order to avoid the necessity of presenting even more technical
details, we present the argument for the case where the slow subspace
contains a single mode, rather than a pair of nearly degenerate modes.
Since $V'$ is symmetric its smallest eigenvalue is given by 
\begin{equation} \label{mp}\lambda_{V'}=\min_{\|u\|=1}
u\cdot(\tau A+{V\over1+\tau V})u\ .\end{equation}
It follows from our assumptions that there exist (normalized)
eigenvectors of $V$
and $A$:
\begin{equation} Ae_A=\lambda_1(1+c_A\epsilon)e_A\ ,\qquad
Ve_V=f(\lambda_1)(1+c_V\epsilon)e_V\ ,\end{equation}
where $\lambda_1$ is the smallest positive eigenvalue of $-\partial^2$
and $e_A$ and $e_V$ are close to $e_1$, the corresponding eigenvector.
The variational principle then gives immediately a lower bound on
$\lambda_{V'}$,
\begin{equation} 
\label{lolambda}\lambda_{V'}>f(\lambda_1)(1+\epsilon c_{V'})\ ,\end{equation}
for some $c_{V'}$ between $c_A$ and $c_V$.
To get an upper bound on $\lambda_{V'}$ recall that it was shown in
sec.~\ref{sec:specta} eq.~(\ref{evc4})
that the $k$ component of $e_A$ is of order
$\epsilon k^{-1}$, and note that the bound~(\ref{bounda}) implies
\begin{equation}\label{boundv} 
f(-a_{\rm min}\partial^2)<V<f(-a_{\rm max}\partial^2)\ .
\end{equation}
We now use $u=e_A$ in eq.~(\ref{mp}) and get
\begin{equation}\label{uplambda}
\lambda_{V'}<\lambda_1(1+c_A\epsilon)+e_A\cdot f(- a_{\rm max}\partial^2)e_A
<f(\lambda_1)(1+\bar c\epsilon)\ ,\end{equation}
for some constant $\bar c$. Eqs.~(\ref{lolambda})
and~(\ref{uplambda}) establish the desired property of the
eigenvalues.

The corresponding eigenvector $e_{V'}$ is the one which minimizes
eq.~(\ref{mp}). Because of the minimax principle applied to $A$ and 
$V$, letting $u=e_1+w$ with $w\cdot e_1=0$ and $w$ small, the
quadratic form 
$u\cdot Au$ may be approximated by

$$ u\cdot Au\sim (w-w_A)\cdot A(w-w_A)+\lambda_A\ ,$$
and similarly
$$ u\cdot Vu\sim (w-w_V)\cdot V(w-w_V)+\lambda_V\ .$$
Therefore, in order to find $w_{V'}$ we need to minimize the quadratic form
$$\tau (w-w_A)\cdot A(w-w_A)+(w-w_V)\cdot {V\over1+\tau V}(w-w_V)\ .$$
The minimum occurs at
\begin{equation} w_{V'}=(1+B)^{-1}
(Bw_A+w_V)\ ,\end{equation}
where
$$ B=\tau{1+\tau V\over V}A \ .$$
We can use again the bounds~(\ref{bounda}) and~(\ref{boundv}) to show
that
$$ b_{\rm min}g(-\partial^2)<B<b_{\rm max}g(-\partial^2)\ ,$$
for some positive numbers $b_{\rm min},\,b_{\rm max}$ and a positive
function $g$, and thus bound the components of $w_{V'}$,
$$ |(w_{V'})_k|<{(w_{V})_k+b_{\rm max}g(|\mu_k|^2)(w_{A})_k
\over 1+b_{\rm min}g(|\mu_k|^2)}\ ,$$
where $|\mu_k|^2$ is the $k$th eigenvalue of $-\partial^2$ (see
sec~\ref{sec:specta}). This shows that $(w_{V'})_k\sim\epsilon k^{-1}$.
This completes the demonstration of the desired properties of $V'$, and,
on applying the results of section~\ref{sec:oneplusa}, the existence
of hydrodynamic Lyapunov modes in the symplectic case.

\section{Numerical tests}\label{sec:num}
The purpose of this section is to verify numerically some of the
statements given above, and on to further study numerically the dependence
of hydrodynamic behavior of several Lyapunov modes as a function of
noise level as well as system size.

The simplest system we discuss is a product $\prod_n A_n$ of
independent matrices of
the form~(\ref{forma}). Since the relative gap between the first two
non-zero eigenvalues is $\OO(1)$ [see
eqs.~(\ref{blockd}--\ref{block2})], the contraction in this case is
strong, and the potential problems of the accumulation of errors
as discussed in
sec.~\ref{sec:oneplusa} and appendix~\ref{ap:counter} are absent. 
Nevertheless, even in this case, there are some qualitative
differences between the behavior of the Lyapunov modes, and the
corresponding eigenmodes of a single matrix.

We quantify the degree of hydrodynamic behavior in the Lyapunov modes
as follows. For the Lyapunov vectors $v_i$ we computed the {\em
residuals}
$r_i$,
that is, the norm of the orthogonal complement 
$$ r_i=\|v_i-(v_i\cdot e_k)e_k-(v_i\cdot e_{-k})e_{-k}\| $$
where $k=k(i)$ is the wave vector associated with vector $i$. (For example,
$k=1$ for the vectors $v_{N-1}$ and $v_{N-2}$ discussed above.)
In fact, to get more precise results we subtracted from $v_i$ all the
components with lower-lying $k$:
$$ r_i=\|v_i-\sum_{k\le k(i)}\bigl((v_i\cdot e_k)e_k-(v_i\cdot
e_{-k})e_{-k}\bigr) \| $$
(The results are not very different for the two definitions
of $r_i$.)

\begin{figure}
\epsfbox{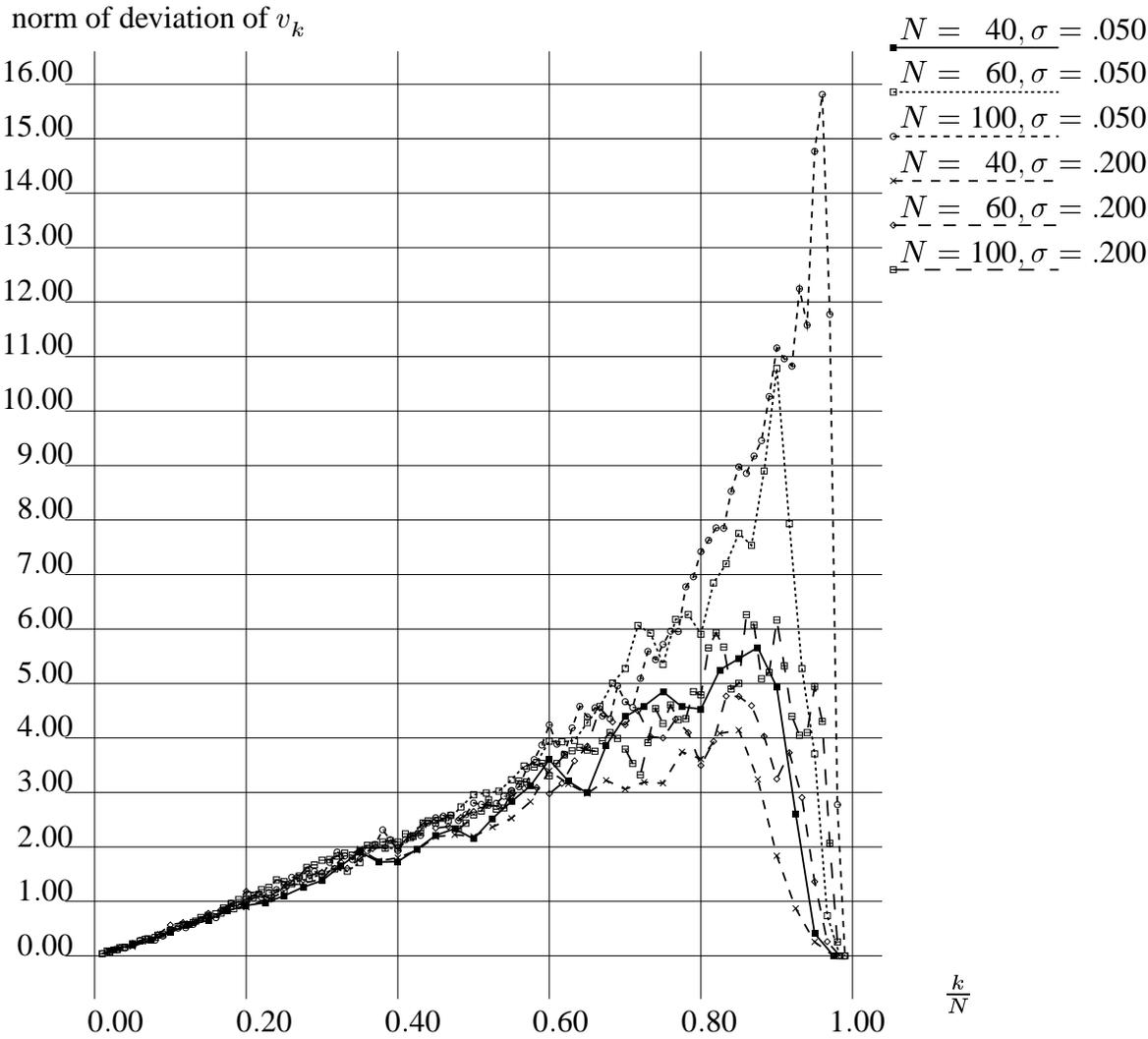}
\caption{\label{fig:0vec}Data collapse of the residuals of the
Lyapunov 
vectors of a product of
random Laplacians. The
curves show $\left({r_k/\sigma}\right)^2$ as a function of
the wave number $k/N$ for several values of $N$ and $\sigma$. 
The data points are
averages of 10 realizations, averaged also within each pair
which corresponds to the same $|k|$. As expected, the range of collapse
increases with the system size. The random variables are chosen as
$a=0.5\pm \sigma$ with probability $1/2$.} \end{figure}

Fig.~\ref{fig:0vec} presents these residuals for systems with
different sizes of the matrices and different values of the noise variance
$\sigma$. The vertical axis measures $(r_k/\sigma)^2$, and the horizontal
axis gives $k/N$.
The approximate collapse of the graphs for small $k$ implies that the
dependence of the residuals on system size $N$ and noise strength
(variance) $\sigma$) is given by the scaling form 
$$r_k=\sigma f_1\left({k\over N}\right)\ .$$
The behavior of $f_1$ for small $x$ is approximately $f_1(x)=\OO(\sqrt x)$,
which implies that for a fixed $k$
 $$r_k\sim {\sigma\over\sqrt N}~,$$ the {\em same} dependence as in
the residuals of the eigenvectors of a single matrix. However, the
dependence of the residuals as a function of 
$k$ is $r_k\sim\sqrt k$, slower the linear dependence on $k$ in
the case of a single matrix.

\begin{figure}
\epsfbox{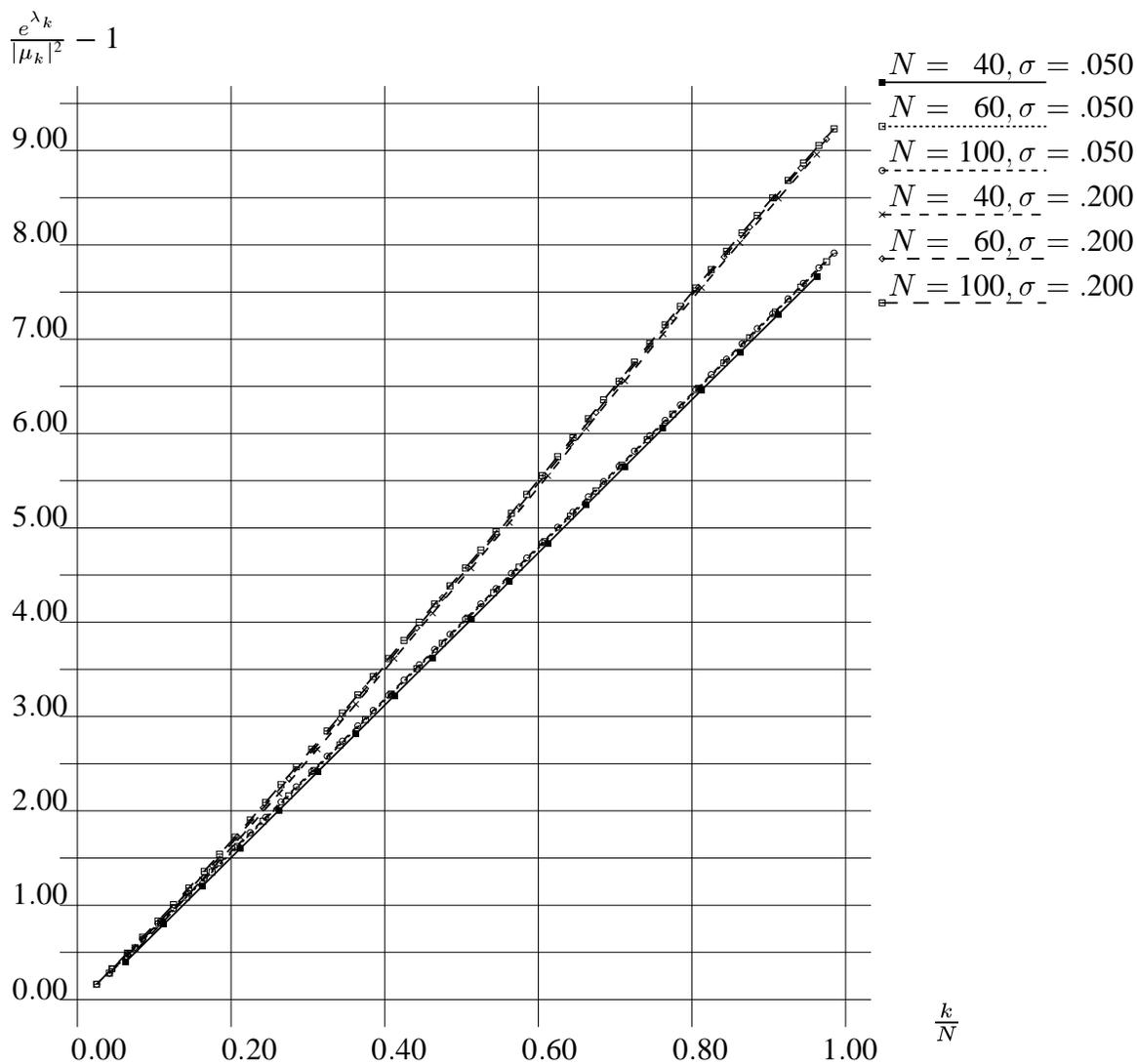}
\caption{\label{fig:0exp}Data collapse of the relative deviations of
the Lyapunov 
exponents of a product of random Laplacians. The 
curves show ${\delta_k/\sigma^2}$ as a function of
the wave number $k$ for several values of $N$ and $\sigma$.
In fact the data collapse is perfect for fixed $\sigma $.} \end{figure}

For
the Lyapunov exponents $\lambda_i$ we measure the relative
deviation $\delta_k$ from the respective eigenvalues $|\mu_k|^2$
of the discrete
Laplacians (see sec.~\ref{sec:specta}),
$$ \delta_k={\exp(\lambda_k)\over|\mu_k|^2}-1\ . $$
The results for the deviations $\delta_i$ of the Lyapunov exponents
are displayed in 
fig.~\ref{fig:0exp}, where
${\delta_k/\sigma^2}$ is plotted against ${k/ N}$.
The data collapse implies that
$$ \delta_k=\sigma^2f_2\left({k\over N}\right)\ .$$
The function $f_2(x)$ is approximately linear for small $x$ which
implies for 
$k\ll N$:
$$\delta_k\sim r_k^2\ .$$
This is not unreasonable, since the Lyapunov
exponents, unlike the eigenvalues of a single matrix, are given as a
result of an averaging process.

\begin{figure}
\epsfbox{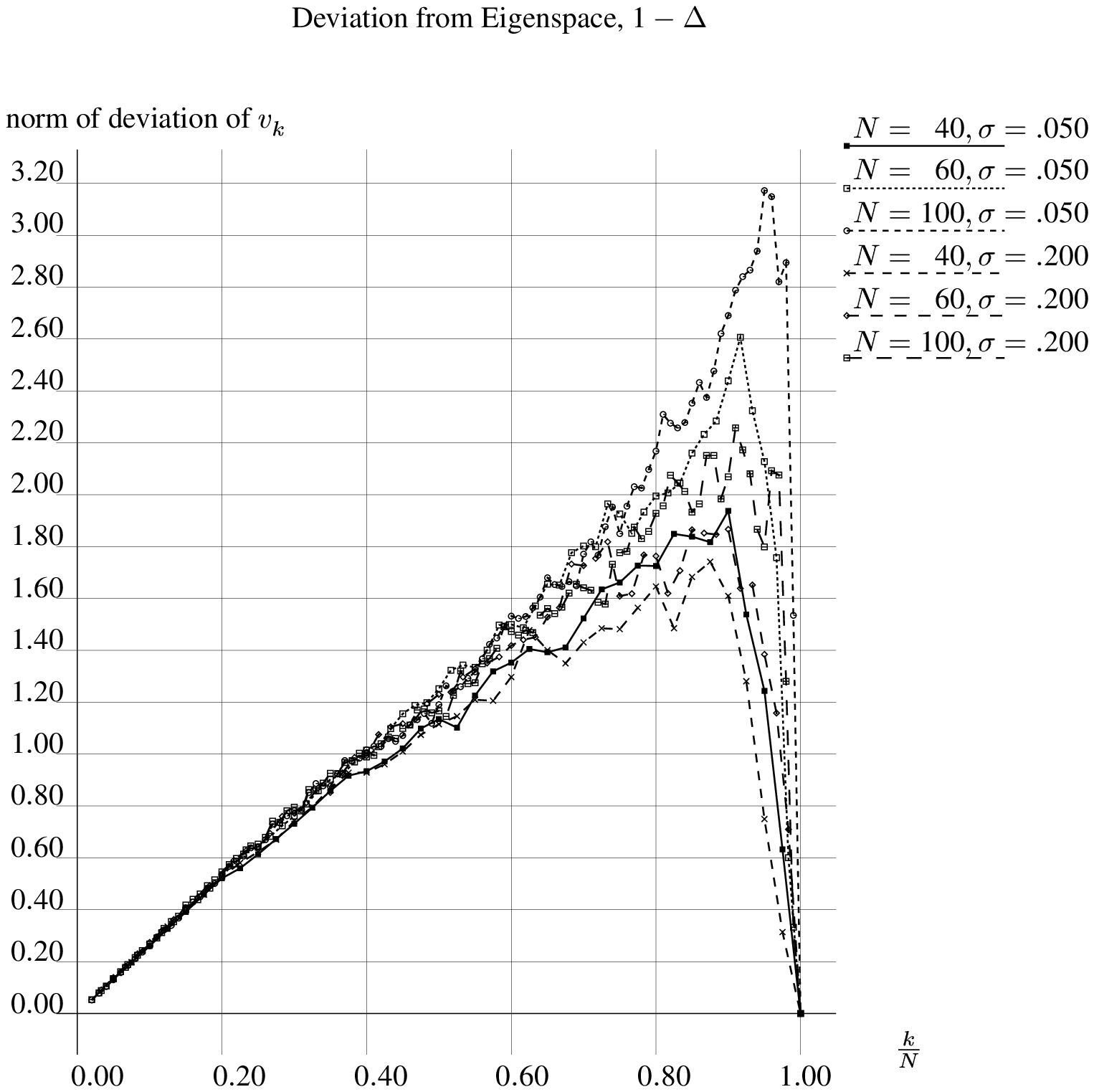}
\caption{\label{fig:1vec}Same as fig.~\ref{fig:0vec} for the product $\prod_n(1+A_n)$, 
except that the vertical axis measures ${r_k/\sigma}$.} 
\end{figure}

\begin{figure}
\epsfbox{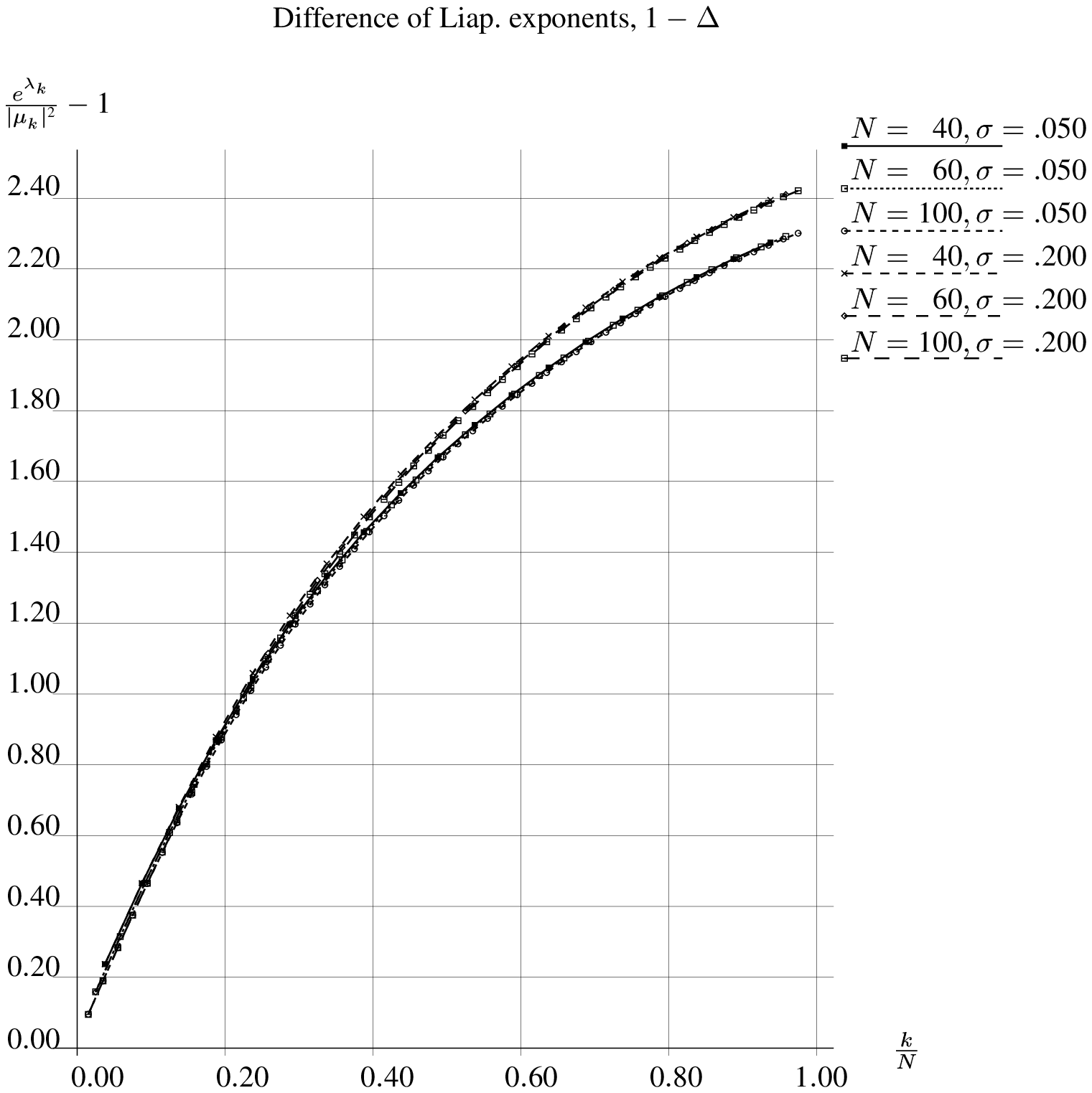}
\caption{\label{fig:1exp}Same as fig.~\ref{fig:0exp} for the product $\prod_n(1+A_n)$, 
except that the vertical axis measures $\sqrt\delta_k/\sigma$.} 
\end{figure}

We next present a similar analysis for the product $\prod_n(1+A_n)$
which was considered in section~\ref{sec:oneplusa}. The results for
the residuals of the Lyapunov vectors and the deviations of the
exponents are presented in figures~\ref{fig:1vec} and~\ref{fig:1exp}
respectively. The scaling form for the residuals is in this case
$$r_k=\sigma f_3\left({k\over N}\right)\ ,$$
where $f_3(x)\sim x$ for small $x$. Thus, in this case the residual
for a fixed $k$ decreases as
$$r_k\sim {\sigma\over N}~,$$ 
that is, {\em faster} than the $N^{-1/2}$ decrease in
the residuals of the eigenvectors of a single matrix. The analysis
presented in section~\ref{sec:oneplusa} is too general to capture this
behavior.


The relative deviations of the exponents scale in this case as
$$ \delta_k=\sigma^2f_4\left({k\over N}\right)\ ,$$
and $f_4(x)\sim x^2$ for small $x$, so that as in the product of
random Laplacians, the relative
size of $r_k$ and $\delta_k$ is
$\delta_k\sim r_k^2\ .$

\appendix

\section{Counter-example}\label{ap:counter}
We want to show here that an `unfortunate' choice of rotations can
move the system out of the region where the Lyapunov vectors remain
essentially aligned with the eigendirections of the Laplacian.
The issue here is that, on one hand, the cones in which these vectors lie are slightly contracted and on the other hand slightly turned. The
`counter-example' shows that the turning wins over the contraction.

Let $V_\f ={\rm Span}\{e_1,\,e_2\}$ and $V_\s ={\rm Span}\{e_3\}$. Suppose
that $L_\f $ contains a vector with block representation
$u=(u_\f ,\,u_\s )$, with $u_\s >0$, normalized such that $\|u_\f \|=1$, and
let $v_\f $ span the orthogonal complement to $u_\f $ in $V_\f $. We construct 
the matrix $\tilde A$ by giving the components in the
representation~(\ref{blockd}),
\begin{equation} R\sim\pmatrix{{\bf 1}&-\epsilon v_\f \cr\epsilon v_\f ^T&1}
\ ,\end{equation}
$D_\s =\alpha\lambda$, and $D_\f $ is such that 
\begin{equation}\begin{array}{c}
D_\f  u_\f =(\lambda+\epsilon^2)u_\f +\epsilon v_\f \\
D_\f  v_\f =\epsilon u_\f +v_\f \end{array}\ .\end{equation}
The image of $u$ is
\begin{equation}
(1+\tilde A)\pmatrix{u_\f \cr u_\s }=\pmatrix{(1+\lambda+\epsilon^2(1-u_\s ))u_\f 
+\epsilon v_\f \cr(1+\alpha\lambda-\epsilon^2)u_\s +\epsilon^2}
\ .\end{equation}
After normalizing the $f$ component to 1, the $s$ component becomes
\begin{equation} u_\s '\sim
\left[1+(\alpha-1)\lambda+\epsilon^2(3/2-u_\s )\right]u_\s +\epsilon^2
\ .\end{equation}
Evidently, even if $u_\s =0$ initially, by choosing $\tilde A$ as above,
$u_\s $ can be increased to an $\OO(1)$ value (as
$\epsilon,\,\lambda\rightarrow 0$) if $\lambda=\OO(\epsilon^2)$.

{\bf Acknowledgments}: We have profited from very useful
discussions with S. Lepri, C. Liverani, Z. Olami, A. Politi, and
L.-S. Young. But we are 
particularly indebted to H. Posch for having provided and discussed
with us his beautiful numerical experiments. This work was partially
supported by the Fonds National Suisse, and part of it was done in the
pleasant atmosphere of the ESI in Vienna and at the Istituto Nazionale
di Ottica in Florence.

\end{document}